\documentclass[12pt]{article}
\usepackage{graphicx}
\usepackage{amssymb,amsmath,amsfonts,palatino,amsthm}
\usepackage{amssymb}
\usepackage{epstopdf}
\usepackage{slashed} 
\newcommand{\f}{\begin{equation}}
\newcommand{\ff}{\end{equation}}

\usepackage{xcolor}
\usepackage[hidelinks]{hyperref}

\begin{document}
\title{Biocosmology: Biology from a cosmological perspective}
\author{Marina Cort\^es,$^{1,2}$ Stuart A.\ Kauffman,$^{3}$ Andrew R.\ Liddle,$^{1,2}$ and Lee Smolin$^1$\\~\\
$^{1}$Perimeter Institute for Theoretical Physics, 31 Caroline Street North, \\ Waterloo, Ontario N2L 2Y5, Canada
\\
$^2$Instituto de Astrof\'{\i}sica e Ci\^{e}ncias do Espa\c{c}o, Universidade de Lisboa,\\
Faculdade de Ci\^{e}ncias, Campo Grande, PT1749-016 Lisboa, Portugal\\
$^3$Institute for Systems Biology, Seattle, WA 98109, USA \\ 
}
\date{\today}
\maketitle
\begin{abstract}
The Universe contains everything that exists, including life. And all that exists, including life, obeys universal physical laws. Do those laws then give adequate foundations for a complete explanation of biological phenomena? We discuss whether and how cosmology and physics must be modified to be able to address certain questions which arise at their intersection with biology.  We show that a universe that contains life, in the form it has on Earth, is in a certain sense radically non-ergodic, in that the vast majority of possible organisms will never be realized.  We argue from this that complete explanations in cosmology require a mixture of reductionist and functional explanations. 
\end{abstract}
\newpage

\tableofcontents


\section{Introduction}

This is the second in a series of papers, starting with Ref.~\cite{biocosmos}, that asks whether and how the kinds of explanations we employ in cosmology must be modified if we are to understand biological organisms as having a natural place in the Universe. How must the role of physics and physical explanations change if we live in a biological universe --- a biocosmos?

The purpose of this article is to present a detailed argument which leads us to a definite answer to those questions. The main argument of this paper is methodological, and it rests on the answers to two questions:
\begin{enumerate} 
\item What kind of explanations are necessary for elucidating biology? and,
\item How can we revise the kinds of explanations we use in cosmology so as to make possible biology as a science of parts of the Universe?
\end{enumerate}

The starting point of our investigation is the affirmation that statistical thermodynamics can be extended to provide the physical basis for explaining how biological systems organize themselves. However, we point out that the existence of life is due to the existence of thermodynamic subsystems whose thermalization times are much greater than the Hubble time, or rather no thermodynamical equilibrium is ever reached while the system is alive and thriving.

These are systems whose descriptions cannot ignore the fact that they are embedded in a dynamical, expanding Universe. An even larger departure from our expectations of how statistical thermodynamics governs physics is the discovery that all living organisms are radically anti-ergodic, in the sense that only a tiny fraction of their state spaces will ever be realized, even if all the baryons in the Universe were to be available for many orders of magnitude of Hubble times. Because of the vast capacity of phase space, these systems will never be in equipartition and will never approach an equality of time and ensemble averages (so long their circumstances remain the same), which is the ergodic hypothesis to be satisfied by standard thermodynamical systems in equilibrium. This arises because the number of types of living organism that could exist, even if the Universe were crammed full of life, is vastly smaller than the number of organisms that could be coded in DNA or RNA or catalyzed and built with proteins. The state space is then extraordinary sparsely occupied. The main question that must be asked concerning any organism or biological molecule is why does it exist, while vastly more alternatives do not.

Such questions have answers, but bottom-up, purely-reductionist, explanations are not enough --- we will never understand life by evolving it from initial conditions, no matter how well we understand the fundamental laws. What is needed to incorporate biology into physics and cosmology is a combination of reductionist and functionalist explanations. Neither alone suffices.

The methodological arguments get their relevance from the conspicuous and suggestive fact that the life emerges and evolves on cosmological timescales. To put this simply, during the roughly 3.8 billion years that life on Earth has taken to evolve to its present state, the Universe has grown and altered significantly. 
 
\section{Three kinds of cosmological statistical systems}
 
We start by affirming that the right framework to situate the study of large macroscopic subsystems of the Universe is statistical thermodynamics \cite{statmech}. Whether we are interested in systems in equilibrium or far from it, in gases or the most complex of systems, we want to formulate descriptions in terms of probability distribution functions on phase space or state space. These provide coarse-grained descriptions of how macroscopic systems evolve under the fundamental laws. Note that when we name the state of equilibrium of a system we need to specify the interactions with regard to which the system is in equilibrium. That is because the equilibrium state of a hot gas looks very different from the equilibrium state of a living system.
 
This leads to our first and fundamental point, which is that if we wish to apply notions from statistical thermodynamics to subsystems of the Universe, we must distinguish three different kinds of systems, which we will call Type I, Type II, and Type III systems. These are defined in terms of the thermalization time $t_{\rm thermal}$, which is the time it takes for the system in question to return to a stable or metastable state of thermal equilibrium after an order unity perturbation.

Additionally, ordinary statistical systems in equilibrium are typically ergodic, i.e.\ they satisfy Boltzmann's ergodic hypothesis, which says that over long periods of time, the portion of time spent in any region of the accessible phase space is proportional to the volume of that region. Closely related to this is equipartition: the postulate that in thermodynamic equilibrium all accessible microstates are equally probable.

Metastable states, in which the system can be trapped in a false ground state with a long decay time $t_{\rm constraints}$, are common in astronomy;  typical examples are red or brown dwarfs, which are the lowest mass stars that undergo nuclear fusion. Red dwarf masses extend down to about 0.08 Solar masses, so their lifetimes extend up to $10^{13}$ years (about a thousand Hubble times). Because they may burn meta-stably for Hubble times or longer, these metastable states play an important role in cosmology. Conversely, cosmology, i.e.\ the fact that they exist in an expanding Universe, may be important for gaining a complete description of stars which live many Hubble times.

Metastable states may occur due to the presence of constraints or high potential energy barriers that either drastically slow down, or prevent altogether, the approach of the system to its true equilibrium state.  An example of the latter situation is a white dwarf star, whose final reserve of stored energy can only be released via accretion or collision with external objects taking it over the Chandrasekhar limit. We will refer to this as a system in (metastable) equilibrium, subject to constraints. 

Systems like these are very common.  There are two relevant timescales. First  $t_{\rm constraints}$ is the timescale over which the constraints hold; on longer timescales they may fail or leak. For these systems we will use  $t_{\rm eq-con}$ to refer to the time it takes to
evolve to equilibrium on the constrained subset of states, i.e.\ the states allowed
by the constraints.  Provided  
\f
t_{\rm eq-con} \ll   t_{\rm constraints}
\ff
a system will, over intermediate time scales $t$ defined by $t_{\rm eq-conl} \ll t \ll t_{\rm constraints}$, behave in some ways like an ordinary equilibrium system. In particular it may, given a sufficiently-long time, be ergodic on the restricted state space created by the constraints. With the same restrictions, it may achieve equipartition. 

An important example of such a system is a white dwarf star, which we will discuss below. 

Cosmological evolution is characterized by the Hubble timescale
\f
t_{\rm H} \equiv H^{-1} \quad \mbox{where} \quad H \equiv \frac{\dot{a}}{a} \,,
\ff
whose present value $t_{\rm H} = 14.5$ billion years is also an excellent proxy for the current age of the Universe, 13.8 billion years. We will also use the timescale of the {\it stellar era}, over which galaxies form stars  which then burn through nuclear fusion, which at about $10^{13}$ years is three orders of magnitude larger than the present Hubble scale \cite{stellarera}.

We can now define the three system types.

\subsection{Type I systems}

We start with ordinary systems in statistical-mechanical equilibrium. We will categorize Type I systems as systems that satisfy Boltzmann's ergodic hypothesis \cite{ergodic} and for which the thermalization time $t_{\rm equil} \ll t_{\rm H}$.

Following Ref.~\cite{ergodic}, an isolated system satisfies the ergodic hypothesis if, for a suffi\-ciently-long timescale $t_{\rm ergodic}$, averages over times longer that $t_{\rm ergodic}$ are equivalent to averages over an ensemble. These ensemble averages are defined by integration of a probability measure over phase space. It is also the case that these averages do not depend on what starting time is taken. 

A system can fail the ergodic hypothesis for several reasons \cite{brokene}. It can fail to be isolated, or there may be a spontaneous breaking of a symmetry. There may also be constraints or barriers, which either lengthen the thermalization time or define metastable states.

The Universe was a Type I system early in its evolution, when the temperature of the radiation component was large compared to nuclear binding energies (together with all other interactions in the Universe, gravity, vacuum energy, etc., which were still switched off at those temperatures) and material was distributed approximately homogeneously. Whilst this was true it existed as a gas or plasma of elementary particles, first quarks and gluons and, after a phase transition, protons and electrons. For the most part equilibrium thermodynamics suffices to describe the statistical physics of the Universe in that era.

\subsection{Type II systems}

The second kind of system is that where the timescale to come to equilibrium, subject to constraints, is significantly larger than the Hubble time, 
\f
t_{\rm eq-con} \gg t_{\rm H}	\,.
\label{e2}					
\ff
That is, Type II systems will eventually come to (possibly metastable) equilibrium, subject to constraints, although they will take longer than the Hubble time to do so. 

One might have thought this is an empty category, but we already pointed out that such systems are common. The category includes many stars and galaxies \cite{LOTC}. Very remarkably, the time for a star to come to equilibrium is vastly larger than one would naively expect from nuclear timescales. This is due to a combination of two circumstances. First, there is a landscape of several hundred stable nuclear bound states, with binding energies in the MeVs. This generates a complicated construction of constraints and barriers to evolution to equilibrium, which results in a large number of metastable states that trap and sequester nuclear potential energy. The second is that gravitationally-bound systems, including stars and galaxies, have negative specific heat. Such systems heat up when energy is removed, so that they do not come to unique or homogeneous equilibrium states.

The reason this is significant is that, after an initial period of cosmic nucleosynthesis when the temperature first falls below that of the binding energy of helium and the existing thermal population of neutrons is incorporated into helium nuclei (plus trace amounts of a few other light elements), the subsequent burning of the remaining hydrogen to helium, and then up through the heavier nuclei, takes place in stars \cite{nuclei-stars}. For most of their lifetimes, stars establish a dynamical equilibrium between the gravitational binding and the nuclear burning; this is sustained by a feedback loop which drastically slows the timescale for the nuclear matter in stars to burn \cite{LOTC}. 

As we described above, most stars, and most of the baryonic mass, never reach the equilibrium specified by their nuclear binding energies. They do reach metastable ground states of the nuclear interaction, which have the properties of equilibrium states when restricted to the states left accessible to them by constraints. For most stars, roughly those lighter than a Solar mass, this implies a time-to-equilibrium several or many times greater than the Hubble time. Indeed a white dwarf left in isolation would be supported by electron degeneracy pressure even after cooling and persevere essentially indefinitely, disrupted ultimately only via quantum tunnelling through nuclear reaction barriers. 

Such systems are but one example of those satisfying Eq.~(\ref{e2}).

Finally we note that large classes of non-equilibrium systems, such as the dissipative systems studied by Prigogine and Nicolis \cite{Prigogine}, those governed by Morowitz's cycle theorem \cite{Morowitz},\footnote{This states that ``In steady-state systems, the flow of energy through the system from a source to a sink will lead to at least one cycle in the system." or more generally that ``The flow of energy through a system acts to organize that system.''} Kauffman's autocatalytic networks \cite{KOoO}, and Turing's model of pattern formation \cite{Turing}, are all examples of Type II systems. Because the flow of negative entropy\footnote{Throughout, this refers to a decrease of entropy of a subsystem via an external flow of energy with very low entropy, e.g.\ the Earth receiving low-entropy photons from the Sun.} through the system is slow, there is time to establish steady states in non-equilibrium systems, and this by itself drives the formation of structures and patterns. These are Type II because, while they are not in equilibrium, they flow to systems which are in equilibrium subject to constraints \cite{LOTC}.

\subsection{Resolving the early Universe thermodynamic paradox}

This subsection bridges the gap between Type II and Type III systems, which we will define in the next section. In particular we will address a point of confusion in the literature regarding states of equilibrium in the early Universe. During the early Universe its energy budget is strongly radiation-dominated, with a distribution very close to homogeneous, and in thermal equilibrium. At that time the Universe can be well described as a Type I system, and as such, as the Universe continues to expand and cool down one would expect that it stays in thermodynamical equilibrium. And yet, after roughly three minutes, the Universe will spontaneously move away from this equilibrium, and ends up finding itself in a region of phase space which is far from it. Under thermodynamical laws this behaviour is impossible if no external agents are at play. As such this spontaneous move away from equilibrium appears to form a paradox that violates the laws of thermodynamics. Let us analyse it in detail.

Equilibrium was established at very high temperatures, and is maintained as the Universe expands and the radiation cools. The Universe then undergoes a phase transition, big bang nucleosynthesis (BBN), when the temperature falls below $T_{\rm nucleo}$, and is cool enough for the strong nuclear force to switch on. This means that, now, stable nuclear fusion of the first nuclei will survive disruption by high-energy photons, and the formation of the first (stable) light elements can take place. A new phase of the Universe's evolution begins here: a very slow progression towards the newly-unveiled state of equilibrium. Whereas before equilibrium meant equipartition of the available energy amongst individual particle species, it now means minimising the nuclear binding energy, which is the interaction that was switched on during BBN. The state of maximum entropy will be (if we ignore gravity for a moment) a universe of stable/long-lived iron nuclei (the most bound atomic nucleus, per nucleon). This will be achieved by the formation of heavier and heavier elements of the periodic table; the lightest elements formed around 1 to 200 seconds after big bang and the heavier ones formed much later in stellar nucleosynthesis.\footnote{Ref.~\cite{BL20} studies how different the laws of physics would have to be for the Universe to emerge from nucleosynthesis directly into the iron-dominated state.} 

So, during BBN the Universe is suddenly disrupted from its previous equilibrium to a state which is very far from equilibrium. This doesn't mean that the entropy goes down during BBN, but rather that a new ceiling of maximum entropy, much higher than the previous, has opened up. Now the Universe will begin its evolution towards this new ceiling of entropy, which it will achieve when the newly-formed elements have been burned in the cores of stars and their end stages to the constrained equilibrium distribution.

It would appear that there is a contradiction here: the Universe was already in thermal equilibrium and had maximal entropy before cosmic nucleosynthesis. How did it then come to find itself, spontaneously and without external influence, so far from equilibrium? How can there, after BBN, occur a boost in the (newly) maximum attainable entropy that will be responsible for the creation of stars, the burning of their fuel in nuclear reactions, and all of the Universe's subsequent evolution until today? In other terms, given that total entropy was already extremized, how could the Universe spontaneously move away from equilibrium? Nucleosynthesis appears to violate the second law, with its prompt, unaided, displacement from equilibrium and the corresponding boost in the maximum available entropy, providing the Universe, which was on its way to a fate of heat death, with a new breath of life. We shall see that no such violation is required for understanding the phenomenon at hand. 

There is a simple answer to this conundrum: the computation that attributes equilibrium and maximal entropy to the pre-BBN universe is incomplete. The pre-BBN universe was indeed at equilibrium, subject to the constraint $T_{\rm nucleo}<T$, but it was an equilibrium state of the high-energy regime of the full theory, the quantum chromodynamics Lagrangian, ${\cal L}_{\rm QCD}$. At those high energies bound states of hadrons, like neutrons and protons, are not possible. The paradox exists only if we don't include all known interactions in our estimation of the entropy. 
 
At low energies these bound states become stable, and the nuclear interaction becomes possible. With these new bound states included in the accounting of the entropy budget of the Universe, the apparently-spontaneous move away from equilibrium ceases to be surprising or paradoxical, and it not only can be explained but is also expected and predicted.\footnote{Plus notice that we are leaving out gravity out of the full Lagrangian for now, since it is still switched off at high temperatures. However, as we will see, it will switch on soon after and will rule the subsequent evolution of the entire Universe during most of its history up to the present time.} After all, had the distribution of matter and radiation been already in the equilibrium of the full theory, it would have remained in that state till today, with the cooling of the Universe leading to its heat death.

We call the pre-BBN universe a `local' equilibrium, in the sense of locality in energy and temperature: once the Universe moves away from this particular temperature, and falls below $T_{\rm nucleo}$ new equilibrium states will be unveiled that are different from the equilibrium state prior to cosmic nucleosynthesis. The probability density of the Universe, which was near equilibrium at high temperatures before nucleosynthesis, has evolved into a region of the state space that was inaccessible beforehand, and entered a small corner, of states which are far from equilibrium and will henceforth evolve towards it. 

However note that, unlike as argued in Refs.~\cite{Carlo,Carloetal}, the states in this region of configuration space are not novel additions to Hilbert space, nor do they constitute an expansion of the volume of configuration space. Rather, they are states that were {\it always} present in the statistical ensemble of the full theory, but were unoccupied, and unavailable to the Universe at high energy, in regimes in which the temperature was too high for their bound states to survive.

This means that when the temperature is above $T_{\rm nucleo}$ the completeness relation of the Hilbert space sums over {\it all} states, both bound nuclei and free. The bound states are part of the completeness relation but unoccupied at high temperature.

One can describe this by saying that the nuclear bound states consistent with the constraints of the fundamental theory were `unveiled' because the system at low temperatures is now in a different effective regime, but they don't expand the Hilbert space of the fundamental interactions (nor that of the Standard Model). 

We call this distinction between equilibrium states in different effective field theories {\it true equilibrium} and {\it false equilibrium} in an analogy to the true and false vacua of field potentials in cosmology and particle physics. Just like the false vacuum of potential energy, the state of false equilibrium at a given temperature is unstable, or metastable. It will continue to evolve towards the true equilibrium, once constraints allow, which in the BBN example happens when $T<T_{\rm nucleo}$. The state of true equilibrium is the state which maximizes entropy in the thermal ensemble in a calculation that includes all known interactions at play in the system.

Most importantly here we want to emphasise that equilibrium and entropy are theoretical concepts and not observable, measurable quantities of Nature. While observables describe the state of the system, entropy describes the observer's {\it knowledge} of the system.\footnote{This is with the exception of well-controlled purely-thermodynamical closed systems where the sole contribution to entropy comes from thermal interactions. Under such particular circumstances entropy can be accurately estimated through measurement via the first law of thermodynamics.} Both the state of equilibrium and the value of entropy of a system result from theoretical deduction, and therefore are dependent on the assumptions going into the derivation. In particular if we leave interactions that are present in the system out of the calculation, we will no longer be able to explain the system's long-term behaviour, as we have seen in the example of the BBN phase transition. This has created the misunderstanding of a paradoxical viewpoint, due to the fact that major interactions at play in the Universe were neglected in the estimation. 

We can summarise this as,
\begin{center}
\textit{The value of entropy calculated for a system is not an observable, and depends on the number and nature of the interactions included in the calculation. An observable is a function of the state of the system, but entropy is a function of the observer's knowledge of the system.}
\end{center}

A good example of how entropy is not a measurable quantity can be found in the cosmologists' evolving understanding of the Universe's entropy, in recent decades. In Ref.~\cite{biocosmos} we present in detail an account of this evolution, following Egan and Lineweaver \cite{EL}. The first estimates of the (observable) Universe's entropy date from around 1980 and stemmed from thermodynamics, attributing the bulk of the Universe's entropy to the blackbody radiation of the cosmic microwave background, $S_{\rm relativistic}\sim 10^{90}$ \cite{KT}, expressed in natural units, $G=c=\hbar=k_B=1$. Some years later the realisation that the entropy of gravity could not be overlooked in the overall budget led to this value being revised by Frautschi \cite{Frautschi}, and later firmed up by Penrose  \cite{Penrose}. The conclusion is that the formation of supermassive black holes has led the gravitational contribution to today dominate over that from the microwave background by around 15 orders of magnitude, $S_{\rm BH}\sim10^{104}$. 

More recently this estimate has had to be revised yet again to account for the fact that the vacuum energy $\Lambda$ dominates the Universe's energy density today, so that the corresponding entropy that $\Lambda$ generates at the cosmic horizon has to be considered. It turns out our previous estimate of the Universe's entropy due to gravity is wrong yet again, and altogether negligible when compared to the entropy of the vacuum which dominates over the gravity value by a further twenty or so orders of magnitude, $S_{\Lambda}\sim 10^{124}$ \cite{EL}. 

The point we are making here, with the example of the evolving understanding of the Universe's entropy,  illustrates our earlier premise that the estimated entropy for a system is a function of the observer's knowledge of the system. Had the entropy of the Universe been an actual observable, these erroneous estimates for total entropy would have been ruled out by observation alone. Instead, each of the values $S_{\rm relativistic}$ and $S_{\rm BH}$ were revised when they were discovered to be subdominant in light of new theoretical knowledge on the composition and the nature of interactions present in the Universe. With regard to today's entropy of the Universe, $S_{\Lambda}$ is our most reliable estimate. However nothing forbids that, should our understanding of the composition of the Universe continue to evolve, further additions can potentially have to be considered, and the entropy value revised once again. 

This subsection conveys the message that, in order to avoid apparent paradoxes, when we estimate the equilibrium state of a system, we need consider, to the best of our knowledge, the single microscopically-complete Hilbert space. It then follows that entropy is maximized or not at a given time depending on constraints intrinsic to the system. 

\subsection{Type III systems}

Both Type I and Type II systems arrive eventually at a stable or metastable equilibrium. Type I are those that take substantially less than the Hubble time, while Type II take at least the Hubble time to come to equilibrium, so that the expansion of the Universe may influence their evolution.

But we are not yet done. For there are interacting thermodynamic systems that seem to {\it never come to equilibrium}, stable or metastable. That is, they do not reach an equilibrium state so long there is a slow steady flow of energy, which, as we shall see, is a requirement for this sort of systems to evolve.\footnote{Our definition includes the caveat `interacting' in order to exclude systems that do not evolve into complex bound states, but fail to ever reach equilibrium simply because their interactions are too weak for the particles to interact any time in the future of an expanding Universe \cite{nonthermal,Dyson-future}. Without this, systems such as gravitational waves would fall under Type III.} These are the Type III systems. How are such systems possible?

As the Universe expands it cools below the threshold, $T_{\rm mol}$, where bound states of atoms become stable. This opens a landscape of possible molecular states so large that the number of actual states that could be realized concretely, making use of the materials on the surface of a planet, or for that matter all the material in the approximately $10^{22}$ solar systems in our observable Universe, is vastly fewer than the number of possible states.

At this point ergodicity and equipartition are simply not possible and equilibrium will never be reached, subject to the presence of a steady flow from the energy reservoir. These systems are then radically non-ergodic, in that the vast bulk of their state spaces will never be explored or realized, even were all the baryons in the Universe available for many orders of magnitude of Hubble times. These systems will never be in equipartition and they will never approach an equality of time and ensemble averages.

Such a system, which is vastly non-ergodic because the Universe cannot in its whole expanse of space and time populate any more than a tiny fraction of the space of states, is what we call a Type III system.  Their state or configuration spaces are so vast that in any fixed region, such as the biosphere of a planet, the number of particles is utterly insufficient to fill out all but a tiny fraction of the possible bound states. 

This implies that, subject to the condition that the system remains in the situation where the number of possible states vastly exceeds the number of realizations, there is no reasonable timescale over which it might be thermalized, provided the slow flow of energy persists. Thus, we can say that there is no thermalization time.

Biological systems, including those of Earth's biosphere, are prime examples of Type III systems, with the Sun providing a slow and steady flow of energy. Even more, biological systems are to date the only example of Type III we know of. Alternatively, and perhaps more appropriately, one could consider the entire Earth--Sun system to be the Type III system, as that combination is then closer to an isolated system for the purposes of satisfying the second law. Even then this combination cannot be considered a fully-isolated system as both Sun and Earth radiate photons, and hence both energy and entropy. Particularly in the case of Type III dynamics we need to take into account that the Earth receives very low entropy radiation from the Sun during the day and radiates higher-entropy photons at night into outer-space. So the inclusion of the Sun makes clear that there is no paradox of perpetual low-entropy cycles in Type III systems on Earth. After all our biosphere would be very short-lived without the Sun.

This does not mean that Type III systems live forever. In the case of the living organisms, the biochemical systems that sustain them, such as the ribosomes and other machinery of the cell, are always subject to thermal fluctuations, as are the delicate chemical bonds themselves. There are many ways that a Type III system can cease to be Type III --- we would say die or that the whole system becomes extinct. If a Type III system dies it later reaches the state of equilibrium of a Type I or Type II system.

In the case of the biosphere, the existence of Type III systems is due to the vast number of possible stable biological molecules. These are template synthesized according to the genetic code, which can represent many more possible molecules than can actually exist. Each is a set of bound states, and hence we have to consider them to add to the dimensionality of Hilbert space. If we arrange the molecules which exist in Nature by atomic weight, we find that there are a large number of small- and medium-sized molecules, made up of the full variety of atoms, according to the laws of chemical bonding. The Universe is thought to be ergodic up to 500 AMU \cite{SJV}. 
 
But there is a huge landscape or configuration space of biological molecules, with vastly more states, which are the polypeptides and the nucleic acids. These only exist because they are made in biological cells, through specialized and exceedingly-complicated processes. But since life exists we have to count them as part of Nature. 

Essentially, all molecules on Earth larger than a certain size are biological.\footnote{There are a few exceptions, which are however molecular products of human inventions that would not occur otherwise such as synthetic polymers; these may also, for present purposes, be classified as biological.} There are a vast number of these due to the fact that they are built as chains of a small number of interchangeable modules. These are the 4 base pairs which are components of RNA and DNA and the 20 standard amino acids, chains of which comprise proteins. 

A typical protein has a few hundred to a few thousand amino acids; this gives a factor of, say, $20^{1000}$ different protein types. This is vastly more than the entire number of different proteins in the biosphere \cite{Kbeyond}. It is also vastly more than the number of baryons in the observable Universe, and indeed greater than the spacetime volume within our causal horizon in Planck units. This latter is the maximum number of bit-operations that could be realized in the parts of the Universe that could causally affect us. 

Each of these proteins can be coded for by a segment of RNA or DNA. Moreover there is no limit to the length of a protein, so there is no upper bound on the dimension of the Hilbert space of this Type III system.   This is the case both for the Hilbert space of potential bound states and if we limit the description to the Hilbert spaces of actual bound states.  The former increases dramatically whenever a falling temperature goes over a threshold; the latter follows, and may increase the dimension of its state space indefinitely, subject only to the availability of resources. 

Of course, the second law has not been suspended when a system is in the non-ergodic Type III regime, but only temporarily postponed, until the low-entropy energy reservoir is exhausted. The second law continues to be perfectly valid without exception when the entire system is considered. For the case of life on the planet, this means including the Sun in the system (which of course cannot be excluded when studying the dynamics of living systems). The random fluctuations of heat and the accumulation of errors mean that a continual application of work is required to make and repair the long chains of nucleic and amino acids that keep the system far from ergodic, and hence Type III. This requires being situated in a non-equilibrium steady-state environment, where continual flows of energy and cycles of materials are available. When they no longer are, the system degenerates, and returns to Type I or II, i.e.\ dies.

That is, it seems that Type III systems require a source of negative entropy near or within them, and these sources may be Type II systems, as is the case of the Sun. 

Living organisms are all Type III systems. In addition, the biosphere as a whole is a vastly non-ergodic Type III system. 

As we discussed, the era of fusion-burning stars is expected to last roughly $10^{13}$ years, after which the stars become cold burnt-out remnants \cite{stellarera}. A Type III system could thus survive at least this long. But, even after all the stars have died there remain stores of nuclear potential energy, for example locked into the white dwarfs. This energy may be released in collisions or mergers of white dwarf  remnants; or as a result of Type I supernovae caused by the inflow of materials onto a white dwarf. Thus, as Dyson \cite{Dyson-future} and others \cite{stellarera} have found, a sufficiently-intelligent and resourceful Type III system might survive in the Universe indefinitely, by making use of the energy and negative entropy radiated by these sources.

\section{How do we explain the phenomena of Type III systems?}

We now discuss the kind of methodologies required to understand Type III systems. Here we will refer mainly to our own planet and its biology, since the biosphere and subsystems within are the only known examples of Type III, though our discussion is intended to be valid within a broader context of Type III phenomenology.

We first affirm that the aim of science is to discover the best explanation for natural phenomena \cite{Leibniz,Deutsch}. By the best we mean the closest to sufficient reason in Leibniz's sense, i.e.\ a good explanation gives us a convincing account as to why the Universe has feature or property $X$, rather than some alternative, $X^\prime$. The better or best explanation does that more or most convincingly.

\subsection{The failure of determinism and reductionism in Type III systems.}

Let us start by supposing that the standard physicalist methodologies based on determinism and reductionism can be applied to the biosphere. Then given some fine-grained initial conditions we should be able to compute, at least in principle, the value of any well-defined physical observable. Let us consider, for every protein $C$, defined as a chain of up to around 1000 amino acids, the number of molecules $N[C]$ present in the biosphere at time $t$. Indeed, as this is a macroscopic quantity we expect that it can be considered a classical observable that has a definite value. So it must be computable from applying the laws of physics to the initial condition. But $N[C]$ is not computable, for several closely-related reasons we now discuss. 

Let us take an even simpler question: for which amino acid sequences $C$ is the number, $N[C]$, larger than, say, a million? We can't predict this either, for the same reasons.

There are $N_{\rm P}= 20^{1000}$ possible proteins, and of these only around $10^9$ are realized at any one time. This is a tiny fraction of those that could exist. An explanation of the occupation numbers of proteins requires us to understand what is special about the tiny fraction of possible proteins that do get to exist.

It is easy to characterize the ones that do exist; these are the proteins that are coded by the DNA of all the living species at the time $t$. That is, a protein exists if it is coded in the DNA of a species that has been successful at finding a niche in which it prospers.
These are very, very few compared to the species that might have existed.

How do we determine which organisms exist? There is an extraordinarily sensitive dependence on initial conditions. As has often been said, if we ran evolution again, after the same time we would have a completely different collection of creatures. How do we know this? There have been five mass extinctions in the past 550 million years. Each is followed by the emergence of novel species, genera, and even families. This is accompanied by a rapid divergence in the proteins that come to be made after each mass extinction. And each time a different mixture of species proliferates. 

Let $\{ P_{\rm poss} \}$ be the set of possible sets of proteins. For every time, $t$, let $P_{\rm now} (t) $ be the member of $\{ P_{\rm poss} \}$ consisting of all protein types represented in the biosphere time $t$. For each of these we have a number of each kind. 

Given $P_{\rm now} (t) $, a complete knowledge of the microscopic laws of physics, and the locations and velocities of all the molecules, could we predict the future value of $P_{\rm now}$ for $\Delta t=1, 100, 10^6, 10^8 $ etc. years in the future? We don't think this is plausible, for several reasons.

First, we should subject our notions of computability to a reality test. It is known that digital computers do not keep a uniform accuracy; even special-purpose integrators show large uncontrollable errors after some time \cite{SPZ1,SPZ2}. Moreover there are limits on the size and speed of computers that would fit on Earth or even in the Universe \cite{limits-cpu}. 

Additionally, one of the drivers of mutation is radioactive and cosmic-ray disruption of DNA which happens by quantum chance, so even if we could integrate from initial conditions specified to arbitrary accuracy, we still wouldn't be able to generate a unique prediction of outcome. So there is no pragmatic --- in other words actual physically-embodied --- computation that is accurate enough to realize the fantasy of integrating the initial conditions of the biosphere forward from initial conditions.

\subsection{The need for functional explanations in Type III systems}

\label{s:functional}

Unlike Type I systems, a complete explanation of a Type III system is not given when we describe the motions of the atoms in the physical phase space. The reason is we have to explain why a tiny fraction of the possible proteins exist in Nature, while most do not. The same goes for nucleic acid sequences and so on. The small fractions that exist have almost identical physical properties as the ones that do not. How are we to explain the differences?

It is good to take a minute here to emphasize that as scientists, we have stronger commitments than we do to the principles of reductionism and determinism. These include the more basic commitment to finding the truth. Each of these might fail and science will go on, so long as we continue to fulfil our quest as scientists which is, as Deutsch reminds us \cite{Deutsch}, to always seek the best explanation for every phenomenon in Nature.

Following Deutsch (and Leibniz \cite{Leibniz}), the best explanations are both sufficient and complete. They suffice to explain why $P$ rather than NOT$-P$, and nothing more is needed to explain either why $P$ or NOT$-P$.

We have no reason to challenge the assumption that the basic laws of physics govern how Type III systems evolve in time. But it is also the case that they do not give an explanation which is sufficient to genuinely explain the fact that certain proteins exist in the biosphere, while most do not.

The problem is that if the laws plus certain initial conditions explain why $P$ exists, but $Q$ does not, there are many very small changes in the initial conditions that, when evolved by the same laws, will lead to the opposite conclusions. This makes it clear that even if we had infinite precision, evolution of the initial conditions by the laws cannot give by themselves either a complete nor a sufficient explanation for the outcomes \cite{SPZ1,SPZ2,Gisin}.

What kinds of explanations suffice? There is only one kind of explanation that can fill the gap: functional explanations \cite{Nagel,functional1,functional2,functional3,Kbeyond}.

What is a functional explanation? Let us consider an example and work towards a general definition. We want to explain why the molecule $P$ exists in the biosphere, while almost no others do. We propose that a sufficient reason has two parts: 1) $P$ is in the class of molecules that can, according to the laws of physics, be made by a cell, 2) $P$ codes for an enzyme that catalyzes or regulates a chemical reaction that contributes to the survival and fitness of a larger organism it is a part of.

Both assertions are necessary. The first is reductionist and "bottom up", the second is "top down" and refers to a function.

A functional explanation for why some process or component $X$ exists, is one where the reason $X$ exists is because $X$ contributes to the survival or well-being of a larger system. What $X$ does to ensure the well-being of $S$ is its function (in the context of $S$). The point is that explanations in terms of the function of a part or a subsystem play a necessary role in explanations in Type III systems, of which biological systems are the most important example \cite{Kbeyond}.

Several different ideas are wrapped up in this definition of a function. A function is a process, carried out in time. We can talk of an organism, a cell or an animal, as a bundle of processes --- many subprocesses coordinated and tied together. The life of a protein, from formation in the ribosome to its disassembly, is a subprocess within an organism that carries out a function which contributes to the success of the whole. 

This is a functional explanation. Note that functional explanations are needed for giving sufficient explanations for many features of Type III systems. This is because only in Type III systems is there such a radical non-ergodicity that much of what has to be explained is why a structure like an enzyme exists, out of vast numbers that don't. 

In each case the explanation is given in terms of the function of the part in a whole system. Consider this question: why are there hearts in the biosphere? Or why is there a particular kind of enzymes? Reductionism plays a role, since some of the explanation for how the properties of the part --- the heart or the enzyme --- contribute to the well-being or survival of the larger organism or system, is because the component is made of atoms which follow the basic laws of physics. 

But the opposite is also part of the explanation; the heart exists because the cat exists. This is what has been called a {\it Kantian Whole} \cite{Kreinvent,Kbeyond,K-EL}.  The point is that explanations in terms of the function of a part or a subsystem play a necessary role in explanations in biological systems. Functional explanations rely on what are called {\it downward causations}. 

Instances of downwards causation, which are all around us, appear to some reductionists to be illicit. But they misunderstand reductionism to mean that the explanatory arrows must always point one way; from the microscopic laws upward to properties of macroscopic systems \cite{downward}. However, in an evolutionary context which gives precise, historical explanations for how functions and the subprocesses that carry them out arise, the common-sense idea of reductionism is perfectly compatible with both functional explanations and downward causation. This is because a necessary aspect of functional explanations is that the larger organism thrives because the function in question contributes to the organism's survival and fitness \cite{KOoO}. We can't emphasize this basic point enough: {\it in biology there is both downwards and upwards causation. Reductionist explanations work hand in hand with functional explanations: both are present, and necessarily so.}

Indeed, stable and metastable far-from-equilibrium structures are stabilized, as shown by Morowitz's theorem \cite{Morowitz}, because the flows of material and energy develop closed loops, around which they cycle, controlled by feedback processes. These cycles, as well as the feedback loops that stabilize them, combine processes which are the result of downwards causation with processes which are the result of upwards causation. 

Now for each such loop, there is a rate of flow around it of energy and the various materials involved. These are concrete examples of new degrees of freedom which emerge in the special circumstances where one has subsystems such as hearts and enzymes that exist because they contribute to the thriving of a larger Kantian whole.

\section{\bf What is different about Type III systems?}

\subsection{Extreme non-ergodicity and its consequences}

How do we define statistical physics for Type III? We easily see that the usual methods of equilibrium statistical mechanics don't apply. There is no ensemble average because anytime you start the system off in even an infinitesimally-different initial state, the evolution through the space of states or configuration space will be radically different. There is no time average, for the same reason. There is no timescale over which a time average will yield repeatable results. The statistical description of a classical system $S$ must be given, as in Types I and II, in a space of possible microscopic configurations ${\cal P}$ in which the configurations ${\cal C}$ live. In Type I there is a static and uniform probability distribution $\rho [{\cal C}]$. In Type II the system evolves towards such an equipartition over a timescale longer than the Hubble time. In Type III the set of possible configurations is so vast, it dwarves the number of actual realizations of these possible configurations. The dynamics is chaotic so that nearby initial conditions diverge. It is not possible to define or predict a single probability distribution function.

The basic issue is that existence is a very rare property of any particular possible configuration or subsystem. As a consequence a new methodology is needed which is centered on a new kind of explanation, which we argued in Subsection~\ref{s:functional}  is {\it functional explanation.} This is required to explain why those very few that exist, do, while many others do not.

The formal definition of a function goes like this. We work in a Darwinian framework in which a species has a fitness, which is proportional to the chance that it is able to reproduce itself. We define function in terms of Kantian Wholes \cite{Kbeyond,K-EL}.

In a Kantian Whole, the Parts exist in the Universe for and by means of the Whole. All living organisms are Kantian Wholes. You exist in the Universe for and by means of your heart, liver, kidney. These parts exist in the Universe for and by means of you, the Whole. The function of a Part is that subset of its causal properties that sustain the Whole. The function of the heart is to pump blood, not to make heart sounds or jiggle fluid in the pericardial sac. The Parts of Kantian Wholes evolve to create ever-new functions that better sustain the Whole, and they are passed along in evolution by propagation of the Whole. 

More precisely, in a Kantian Whole:
\begin{enumerate}
\item The occupied states or configurations of a Type III system may be characterized as follows: they contain inter-connected bundles of subprocesses, ${\cal K}$, which are connected in the sense that many of the outputs of the subprocesses go into inputs of other subprocesses.

\item These subprocesses can perform tasks and functions.

\item A task is a change in the states of subprocesses which requires work, chemical and/or physical.

\item A function is a more general alteration of ${\cal K}$, its subprocesses or environment, that contributes to the fitness of the Whole, ${\cal K}$.

\item A bundle of subprocesses ${\cal K}$ is then called a {\it Kantian Whole}, in the case that many of its sub-processes, ${\cal P}$, perform functions or tasks that increase or contribute to the fitness of the Whole ${\cal K}$. Each of these ${\cal P}$ exists because it is part of, and contributes to, the fitness of ${\cal K}$, which, due to the contributions of all its subprocesses, thrives. Thus there is a symbiosis among the many subprocesses, $P_I$ (with $I= 1, 2, \ldots$) that make up ${\cal K}$, which exist because by their combined contributions they make ${\cal K}$ viable.

\item The subprocesses exist in the Universe because they make the Kantian Whole viable in its world.  The Parts exist in the Universe for and by means of the Whole.
\end{enumerate}

Kantian Wholes can be arranged hierarchically. A given subprocess may be part of more than one Kantian Whole at the same time, and the Kantian Whole it is part of may be different at different times.

The existence of Kantian Wholes implies a radical change in how we understand and explain Nature. Once we have a Kantian Whole it makes sense to say that a process, event or alteration of its environment, is ``good for $\cal K$", which means ``increases ${\cal K}$'s fitness or survival." 

\subsection{Type III systems can be recursive or excursive and also universal}

Note that a Type III system can realize a function which is recursive, or self-referential. That means it refers to the larger Kantian Whole, $\cal K$, to perform a task that increases the fitness of the Whole, $\cal K$.

It does this because it has a code that provides a complete mapping from all possible chains of nucleic acids to all possible catalytic functionalities. To see why, note that there has been estimated to be roughly $10^8$ different shapes or catalytic tasks which a set of enzymes would need to cover to be universal \cite{KOoO}. 

A coding is a process that allows $\cal K$ to store information about its own structure. This involves an {\it abstraction}, in the sense that a sequence of letters made of nucleic acids on a strand of DNA can represent a sequence of amino acids. A code is {\it an abstraction} if the function that reads and writes information to a subcomponent involves an arbitrary map, which gives a representation of, but is not an instantiation of, that structure.

At this point something very interesting happens. The same coding that gives the information needed to construct itself can be altered to code any of a vast space of alternatives. That is, the organism in its construction explicitly refers to itself (through the coding) and implicitly it refers to a vast set of adjacent possibilities. The first is {\it recursive}, we call the second {\it excursive}.

\subsection{The three arrows of dynamics}

If one takes a Type I system a bit out of equilibrium and lets it evolve from there, an arrow of dynamics emerges which reflects that it is more likely that entropy will increase until the system returns to equilibrium, after which the system fluctuates around equilibrium and there is no longer an arrow of dynamics. Nor was the arrow of dynamics, while it existed, coupled to cosmological degrees of freedom.

We will call this a Type I arrow of dynamics. It is simply the manifestation of the tautology that the most likely outcome of evolution of a closed system is to one of the most probable macro-state. As Paul and Tatiana Ehrenfest showed in 1909 \cite{PT1909} this statement is time-reversal symmetric;  it is equally likely for the entropy to have been higher in the past as it will be to be higher in the future.

In Type II systems, the return to equilibrium can take cosmological timescales, for the reasons discussed above. The result is an arrow of dynamics that is present from the Big Bang to at least three orders of magnitude past the present Hubble time. This second arrow of dynamics might be defined in terms of the lifetime of the Sun or a red or brown dwarf.

But the Type III systems in the Universe have the property that they don't reach equilibrium, subject to the existence of a source of negative entropy such as the Sun. Hence their search for the most probable state defines a third kind of arrow dynamics that can continue indefinitely, or at least for as long as they can find resources. 

The existence of an arrow of dynamics that is not reversible by a finite process raises some interesting questions. Is this compatible with fundamental laws that are reversible, or does this require that the fundamental laws are time-irreversible?

Finally, we briefly note there is in Nature another system which would have fallen under Type III, in the sense that it never approaches equilibrium or equipartition and hence is highly non-ergodic, were it not for our caveat restricting Type III to interacting systems. This is gravitational radiation \cite{nonthermal}, both classical and quantum.  However, a defining feature of Type III systems, and their main reason of interest, is their thriving in dynamical cycles within pockets of phase space with very low entropy, the goal being the study of the complexity that results from that. Gravitational waves do not satisfy these criteria since they do not form complex bound states as living systems do, hence are less attractive from the point of view of this study.\footnote{Cosmic photons and neutrinos also behave as non-interacting systems after their respective decouplings, in that any subsequent energy injection never re-thermalises, though they differ from gravitational waves in having previously been in thermal equilibrium with other particles.}

\section{Proposal for a fourth law of thermodynamics}

\label{s:fourth}

There have been several attempts to formulate a fourth law of thermodynamics, which would characterize how and why steady-state far-from-equilibrium states are able to self organize, in a way that at first look suggests that the second law is being transcended. These attempts were formulated mainly for what we call here Type II systems. From the perspective we present here, the irreversibility in these systems is not due to a transcendence of the second law, but rather to a channelling of negative-entropy flows to delay, sometimes by very long times (as long the slow steady source of energy is active), the approach to equilibrium.

We would like, however, to propose that Type III systems offer a new perspective on irreversibility, which opens up the possibility for a fourth law which is not a re-channeling of the second. 

Let us consider a Type III system, $\cal T$, perhaps a biological organism or the whole biosphere. As discussed above, a Type III system has subsystems which carry out a number of functions for $\cal T$. Let $F_{\rm A}$ stand for the number of {\it distinct, actual} functions being used by $\cal T$ at a given time. We want to distinguish the actual functions from the {\it set of possible} functions, which includes those functions that $\cal T$ already has the possibility of expressing by presently-unexpressed codons, or which could be created in a few steps of mutations or transmutations of the DNA. $F_{\rm P}$ is the number of possible functions which the system might make use of.

Define 
\f
R= \frac{F_{\rm P}}{F_{\rm A}}
\ff
The fourth law we conjecture states that for any Type III system, as long as the non-equilibrium conditions persist which allows the Type III system to maintain its main functions (see below), then
\f
F_{\rm P} , F_{\rm A} \mbox{ and} \; R \mbox{ all have a tendency to increase.}
\ff

In fact, in the one case of Type III systems we have so far treated numerically \cite{biocosmos}, we have found that the ratio $R$ does not merely {\it increase}. It does so at such an unprecedented rate of growth that we needed to resort to the mathematical operation of tetration (involving, at each step, an exponentiation of the argument of the function) to be able to describe the numbers encountered for the increase of phase space, or in this case $F_{\rm P}$. While both increase in Type III systems, the difference between $F_{\rm P}$ and $F_{\rm A}$ is staggering for the example we studied. 

We want to emphasise here that the fourth law is by no means a violation of the second law. If we include the source and outflow of energy as part of the system, as needs be for a consistent treatment, it becomes clear that the total entropy of the system is {\it always increasing}. What Type III systems are doing is rather exploring pockets of phase space of exceptionally-low entropy, which they can reach via the exchange of energy with exceptionally-low entropy from the energy source. By exploiting this almost unlimited source of low entropy --- e.g.\ the Sun --- Type III systems can persist almost indefinitely in a suspension of a low-entropy cycle, subject to the availability of that source. 

\section{A definition of life}

We consider how a definition of Type III systems may augment our familiar definitions of life. These familiar definitions include ``an open non-equilibrium thermodynamic molecular system capable of self-reproduction and evolution''. This minimal definition is often augmented by the concepts of: i.\ Replicating genetic material, ii.\ A metabolism, iii.\ A bounding membrane. Further augmentations include that of thermodynamic work and agency \cite{Kinvestigations}, and constraint closure \cite{MMMM} by which cells do thermodynamic work to construct themselves. 

These features ensure the existence of Kantian Wholes that can evolve and construct ever-new functions in the Universe. The construction of such novelty requires the non-ergodicity extant in Type III systems. 

Note next that the maintenance of a Kantian Whole requires functions which support identity, such as controlling the exchanges of materials as well as the ion concentrations on each side of its membrane. Since these involve moving charges or other molecules against gradients, each of these functions requires work. Hence there must be a flow of energy through the system.  

Work is the constrained release of energy into a few degrees of freedom \cite{Atkins}.  A cannon, cannon ball, and exploding powder are examples. The cannon is the constraint, the boundary condition. No constraint, no work. But work can construct constraints \cite{Kinvestigations}. In a constraint-closed system \cite{MMMM}, the set of constraints constrain the release of energy in a set of non-equilibrium processes that construct the very same set of constraints. We construct our artifacts, while cells construct themselves. By heritable variation and natural selection, cells continually evolve to construct boundary conditions that are new to the Universe \cite{Kbeyond,K-EL}.  The new boundary conditions create new phase spaces, and hence new possibilities in the evolution of the Universe \cite{Kbeyond,K-EL}. This is the behavior of evolving life in Type III systems.

Then our definition of life is:
\begin{quotation}
\noindent
{\it A living organism is a Kantian Whole within a Type III system that is a non-equilbrium self-reproducing system with a metabolism,  an identity and a boundary, and which is capable of open-ended evolution by heritable variation and selection or drift.}
\end{quotation}

\section{Summary}

Ever since Schr\"odinger raised the question of {\it What is life?}\ \cite{Sch-WIL} it has been evident that statistical thermodynamics is the right setting from which to approach an understanding of life in the context of fundamental laws and principles. It has long been suspected that non-equilibrium statistical mechanics should provide a general understanding of pattern formation and self-organized systems, which would underlie biology \cite{Prigogine,Morowitz,KOoO,Turing}. 

Thermodynamics also plays a central role in cosmology. It is then natural to approach the question of why the Universe is hospitable to life using the tools and concepts of thermodynamics.  Here we uncover some conceptual and physical distinctions that turn out to be crucial if we want to discuss how biology fits into cosmology from a physics perspective.

The first is that we have to distinguish three kinds of statistical systems. The first two come eventually to thermal equilibrium, but we distinguish them depending on the ratio of the thermalization time to the Hubble timescale:
\f
r=t_{\rm thermal} H
\ff
We know how this plays a role in nucleosynthesis, but it turns out to have a more general significance.

At one end of the scale, for small $r$, are the ordinary equilibrium thermodynamic systems, which satisfy the familiar principles of equipartition and the equality of time and ensemble averages. These apply universally on small scales as well as in the early Universe, when the temperature is so high that it dominates over the other mass and energy scales. We call these Type I systems. They have no coupling to cosmology because the time required by them to thermalize is much less than the Hubble scale.

Type II systems are those whose thermalization times are on the order of or larger than the Hubble time. This is the intermediate case for which 
\f
r \gtrsim 1 \,.
\ff
These are surprisingly common and include many of the main-sequence stars and disk galaxies. These have been somewhat well studied and their role in setting up steady-state non-equilibrium sources of negative entropy that make life possible have been much discussed, as summarised in Ref.~\cite{LOTC}. 

Lastly, here we introduced a new third class of statistical-mechanical system, which we call Type III systems. These are cases for which the rate of expansion and addition of novel states to Hilbert space is so vast and explosive that we cannot foresee the expression of all the states in that Hilbert space within a time of order of the finite lifetime of the Universe. An alternate characterization of a Type III system is that it has at least one subsystem or component, $Q$, that comes in a vast number of alternative versions, $N_Q$, which are continually added onto, and are roughly equivalent energetically, and where
\f
N_Q \gg N_{\rm p} .
\label{NO1}
\ff
where $N_{\rm p} \sim 10^{80}$ is the number of protons within the Hubble distance. This implies a startling difference from equilibrium systems, of Type I, for which the equality of time and ensemble averages quickly obtains.

As a consequence of Eq.~(\ref{NO1}), in Type III systems there is not enough time, space, nor material in the Universe for more than a tiny fraction of the allowed possible states of those subsystems ever to be realized. There are such a vast number of states within narrow energy ranges that equipartition or equality of time and ensemble averages will never be achieved or relevant, while the system is still of Type III, that is, while it still `alive'. If the system dies it will cease to be of Type III and becomes Type I or II and will also most likely `die' to a state of equilibrium. The recognition that Type III systems are very different than others thermodynamically is essential for understanding why there is life in the Universe.

An important example is the possible proteins, which number something like 
\f
N_{\rm protein} \sim 20^{1000} \gg N_{\rm p} \sim 10^{80} \,.
\ff
Only a small fraction, perhaps $10^9$ of these, are realized in our biosphere 

It is even the case that the number of possible proteins is much greater than $10^{120} $ so even at a rate of expression of one protein per Planck volume per Planck time, the vast majority of proteins would never be expressed.

Thus, {\it simply existing} is a {very rare property} of the states of these subsystems. Consequently, as we discussed in detail above, we require {\bf an adequate explanation of why some states do exist, while almost all the others do not.} As we explained, the narrow pure-reductionist style of explanation does not suffice to give a complete explanation to that type of question. One problem is that the effects of noise, thermal and otherwise, would overwhelm any attempt to answer this kind of question by integrating initial data, which is itself going to be noisy. If we continue to search, nonetheless, for a better account of such a  {\it why me and not them?} kind of question, it must involve a new kind of explanation. We proposed above that this must be {\it a mix of reductionist and functional explanations.}

Very briefly, in the context of Darwinian natural selection, a functional explanation as to why some particular molecule or protein, $\alpha$-bazoodle, exists in the biosphere, is that it exists partly because it contributes to the viability or fitness of a larger system $\cal K$, of which it is a part. In Ref.~\cite{biocosmos} we characterised this condition of viability for existence with the expression \textit{``The name of the game is getting to exist''}. Our $\alpha$-bazoodle does this by performing a function, which is a sequence of operations during which some chemical or mechanical work is done on $\cal K$ or its environment, the result of which is to increase the viability or fitness of $\cal K$. Of course, the explanation for the existence of $\alpha$-bazoodle, must also include an argument that the laws of physics are compatible with its both existing and performing the function it does for $\cal K$. Consequently, {\it any adequate explanation for the existence of a sub-system or sub-process in a Type III system must involve a combination of bottom-up reductionist explanations and top-down functional explanations.}

To spell this out:
\begin{itemize}
\item{} Functional explanations are necessary to give a satisfactory explanation for many questions in Type III systems including our biosphere. This is especially true when we wish to understand living systems from a physics point of view, which we began to do in the research programme we launched in Ref.~\cite{biocosmos} and continue here.
\item{} Functional explanations rely on knowledge of the basic physical laws that govern the components and processes of an organism. In this sense, functional explanations are compatible with physicalism.
\item{} But functional explanations answer questions that purely reductionist explanations cannot, because they rely on downward causation, e.g.\ ``hearts exist because they perform a function of pumping blood, which is necessary for animal life."
\item{} Thus, functional explanations, which rely on specific instances of downward causation, complete explanations that na\"{\i}ve reductionism leaves incomplete.
\item{} Living things, and the biosphere as a whole, are sustained by stable cycles of energy and materials, controlled by feedback. These cycles combine functions which express downwards causation with ordinary upwards causation. 
\end{itemize}

Thus, we can say that the reductionist account is valid, but does not give a complete explanation of how and why these systems come to exist. We may say the same for the purely-functional explanation. Both are needed and indeed, both are true. Nor is this a case of complementarity, for they are perfectly consistent with each other. We provide a complete summary of the arguments in this paper in Table~\ref{t:summary}.

\begin{table}
\small
\begin{center}
\begin{tabular}{ | p { 3.35 cm} || p { 3.35 cm} | p { 4 cm} | p {4 cm}|} 
 \hline 
 &\qquad \quad Type I & \qquad \quad Type II & \qquad \quad Type III \\ 
 \hline \hline
 Ergodicity: & Ergodic. & Ergodic and well mixed on times of order the Hubble time and only when taking into account constraints. & Highly non-ergodic, never returns, to exist is rare, over arbitrarily long times only a very small portion of the state space is sampled. \\
 \hline
 Methodology: & Reductionism and upward causation are sufficient. & Reductionism and upwards causation have to be supplemented by downwards causation because constraints and channels lead to feedback and cycles. & We must admit functional explanation, which is a form of downward causation. This goes hand in hand with a weak form of reductionism.\\ 
 \hline
 Ensemble type: & All of state or Hilbert space. & Dominated by capture by limit cycles. & Dominated by capture by Kantian Wholes.\\ 
 \hline
 Sensitivity to small changes in initial conditions: & None, quickly averages away. & Moderate. & Extreme, can result in a very different biosphere or none at all.\\
 \hline
 Equality of time and ensemble averages respecting constraints: & Yes. & Only on much greater than Hubble times. & Never. \\ 
 \hline
 Correlations of fluctuations: & Gaussian. & Gaussian over very long timescales, and subject to constraints.  & Small fluctuations can produce arbitrarily large deviations. \\ 
 \hline
 Emergent phenomena: & None. & Steady-state non-equilibrium systems, dissipative structures, cycles (Morowitz cycle theorem), limit cycles. & Kantian Wholes, biological cells, the biosphere.\\
 \hline
Characterized by: & Boltzmann, Maxwell, Gibbs. & Prigogine, Morowitz, etc. & Darwin.\\ 
 \hline
 What we count: & All possible microstates, consistent with macrostate. & All possible microstates, consistent with macrostate and constraints. & Possible novel functions of possible Kantian Wholes. \\
 \hline 
How to count states: & Equilibrium thermodynamics. & Non-equilibrium thermodynamics, subject to constraints. & Use TAP  equation to count functions. \\
\hline
\end{tabular}
\end{center}
\caption{\label{t:summary}  {\small The complete argument. The TAP (`Theory of the Adjacent Possible') equation \cite{HSK1,OurTAP} is a model for the growth of systems that are defined by functions, either intrinsic or extrinsic, and can grow by combinations, such as  inventions or patents.  It is described in detail in Ref.~\cite{biocosmos}.} }
\end{table}

We turn, briefly, to the payoffs for science of adopting this kind of methodology, which gives functional explanations a necessary role alongside physicalist explanations.

The first payoff is a new kind of definition of life. We introduced it briefly here and will return to it in future work. We believe that no purely reductionist characterization of life will succeed and that every useful characterization of life will involve a mixture of physicalist and functional criteria.

The second payoff is a proposal for a fourth law of thermodynamics, which we presented briefly in Section~\ref{s:fourth}. 

If biology requires this hybrid of reductionist and functional explanations to provide adequate explanations from a physics perspective, and if the science of cosmology is going to fully explain all that goes on in the Universe thus including biology --- an assertion we made in Ref.~\cite{biocosmos} --- then this kind of mixed functionalist explanation must be part of the methodology of cosmology. Reductionism is required to understand the microscopic degrees of freedom of living Type III systems and functional explanation are required to explain the complex, structured degrees of freedom.

Some readers may reply that the argument we are tracing here can be disrupted by assuming that the multiverse exists, and that within it, there are enough copies of a Type III system that we can apply to it a usual microcanonical argument. We could dispute this, but we don't need to, because this argument just proves our point that the kind of methodology and explanation we can apply to such systems depends on assumptions we make about the Universe at large. In other words, there is a coupling between cosmology and biology.

The existence in the Universe of systems whose thermalization times, even when subject to constraints, are much larger than the Hubble time, is going to have to be significant for the thermodynamics and evolution of the Universe. And the full description of such systems must take into account cosmology, because they exist for timescales on which the Universe evolves significantly.  Thus we see there is a need for a hybrid of biology and cosmology, which we call {\bf biocosmology}. 

\section*{Acknowledgements}
 
We are grateful to Stephon Alexander, Barbara Drossel, George Ellis, Carlo Rovelli, and Simon Portegies Zwart for helpful discussions. This research was supported in part by Perimeter Institute for Theoretical Physics. Research at Perimeter Institute is supported by the Government of Canada through Industry Canada and by the Province of Ontario through the Ministry of Research and Innovation. This research was also partly supported by grants from NSERC and FQXi. This work was supported by the Funda\c{c}\~{a}o para a Ci\^encia e a Tecnologia (FCT) through the research grants UIDB/04434/2020 and UIDP/04434/2020. M.C.\ acknowledges support from the FCT through the grant\\ SFRH/BPD/111010/2015 and the Investigador FCT Contract No.\ CEECIND/02581/2018 and POPH/FSE (EC). A.R.L.\ acknowledges support from the FCT through the Investigador FCT Contract No.\ CEECIND/02854/2017 and POPH/FSE (EC). M.C.\ and A.R.L.\ are supported by the FCT through the research project EXPL/FIS-AST/1418/2021. We are especially thankful to the John Templeton Foundation for their generous support of this project.


\begin{thebibliography}{99}

\bibitem{biocosmos} M. Cort\^es, S. A. Kauffman, A. R. Liddle, and L. Smolin,
{\it Biocosmology: Towards the birth of a new science}, submitted simultaneously.

\bibitem{statmech} S.-K. Ma, {\it Statistical mechanics}, World Scientific, Singapore (1985).

\bibitem{stellarera} F. C. Adams and G. Laughlin, {\it   A dying universe: the long-term fate and evolution of astrophysical objects}. Rev. Mod. Phys. {\bf 69}, 337 (2008) [arXiv:astro-ph/9701131]. 

\bibitem{ergodic} L. Boltzmann. See part 2 of Vorlesungen ber Gastheorie. Leipzig: J. A. Barth. 1898. OCLC 01712811. ('Ergoden' on p.89 in the 1923 reprint.) See also wikipedia: \url{https://en.wikipedia.org/wiki/Ergodic_hypothesis}.

\bibitem{brokene} R. G. Palmer, {\it Broken ergodicity}, Adv. Phys. {\bf 31}, 669 (1982). 

\bibitem{LOTC} L. Smolin, {\it Life of the Cosmos}, Oxford University Press, Oxford (1997).

\bibitem{nuclei-stars} R. N. Boyd, {\it An Introduction to Nuclear Astrophysics}, University of Chicago Press, Chicago (2008).

\bibitem{Prigogine} I. Prigogine and G. Nicolis,
{\it Self-Organization in Non-Equilibrium Systems}, Wiley, Chichester (1997).

\bibitem{Morowitz} H. Morowitz, {\it Energy Flow in Biology},
Academic Press (1968).

\bibitem{KOoO} S. A. Kauffman, {\it Origins of Order}, Oxford University Press, Oxford (1993).

\bibitem{Turing} A. Turing, {\it The Chemical Basis of Morphogenesis}, Phil. Trans. Roy. Soc. London {\bf B237}, 37 (1952). 

\bibitem{BL20} L. A. Barnes and G. F. Lewis, {\it Under an Iron Sky: On the Entropy at the Start of the Universe}, Pub. Astron. Soc. Aus. {\bf 38}, E061 (2021) [arXiv:2012.06975 [astro-ph.CO]].

\bibitem{Carlo} C. Rovelli,   {\it  Where was past low-entropy?},  Entropy {\bf 21}, 466 (2019) [arXiv:1812.03578 [gr-qc]].

\bibitem{Carloetal} K. Jeffery, R. Pollack, and C. Rovelli, {\it On the statistical mechanics of life: Schr\"odinger revisited}, Entropy {\bf 21}, 1211 (2019) [arXiv:1908.08374 [physics.bio-ph]].

\bibitem{EL} C. A. Egan and C. H. Lineweaver, {\it A larger estimate of the entropy of the Universe}, Astrophys. J. {\bf 710}, 1825 (2010) [arXiv:0909.3983 [astro-ph]].

\bibitem{KT} E. W. Kolb and M. S. Turner, {\it The Early Universe}, Nature {\bf 294}, 521 (1981); E. W Kolb and M. S. Turner, {\it The Early Universe}, Addison--Wesley, Redwood City (1990).

\bibitem{Frautschi} S. Frautschi, {\it Entropy in an Expanding Universe}, Science {\bf 217}, 593 (1982).

\bibitem{Penrose} R. Penrose,  in {\it General Relativity, an Einstein Centenary Survey}, eds S. W. Hawking and W. Israel, Cambridge University Press, Cambridge (1979);  R. Penrose, {\it The road to reality: a complete guide to the laws of the universe}, Jonathan Cape, London, UK (2004).

\bibitem{nonthermal} L. Smolin, {\it The thermodynamics of gravitational radiation}, Gravity Research Foundation award essay, second prize 1983, Gen. Rel. Grav. {\bf 16}, 205 (1984); L. Smolin, {\it On the intrinsic entropy of the gravitational field}, Gen. Rel.  Grav. {\bf 17}, 417 (1985).

\bibitem{Dyson-future} F. J. Dyson, {\it Time without end: Physics and biology in an open universe}, Rev. Mod. Phys. {\bf 51}, 447 (1979); see also F. J. Dyson, {\it Disturbing the Universe}, Joanna Cotler Books (1979).

\bibitem{SJV} S. A. Kauffman, D. P. Jelenfi, and G. Vattay, {\it Theory of Chemical Evolution of Molecule Compositions in the Universe, in the Miller--Urey Experiment and the Mass Distribution of Interstellar and Intergalactic Molecules}, J. Theor. Biol. {\bf 486}, 110097 (2020) [arXiv:1806.06716 [q-bio.PE]].

\bibitem{Kbeyond} S. A. Kauffman, {\it A World Beyond Physics: The Emergence and Evolution of Life?}, Oxford University Press, Oxford (2019).

\bibitem{Leibniz} N. Rescher, {\it G. W. Leibniz's Monadology}, University of Pittsburgh Press (1991).

\bibitem{Deutsch} D. Deutsch, {\it  The beginning of infinity : explanations that transform the world}, Allen Lane (2011).

\bibitem{SPZ1} T. Boekholt, S. Portegies Zwart, and M. Valtonen, {\it Gargantuan chaotic gravitational three-body systems and their irreversibility to the Planck length}, Mon. Not. Roy. Astr. Soc. {\bf 493}, 3932 (2020) [arXiv:2002.04029  [astro-ph.IM]].

\bibitem{SPZ2} S. Portegies Zwart, {\it Computational astrophysics for the future}, Science {\bf 361}, 979 (2018).

\bibitem{limits-cpu} S. Lloyd,   {\it Ultimate physical limits to computation}, Nature {\bf 406}, 1047 (2000) [arXiv:quant-ph/9908043].
	
\bibitem{Gisin} F. de Santo and N. Gisin, {\it Physics without determinism: Alternative interpretations of classical physics}, Phys. Rev. A{\bf 100}, 062107 (2019) [arXiv:1909.03697 [quant-ph]]; N. Gisin, {\it Mathematical languages shape our understanding of time in physics}, Nature Physics {\bf 16}, 114 (2020) [arXiv:2002.01653 [physics.hist-ph]].

\bibitem{Kreinvent} S. A. Kauffman, {\it Reinventing the Sacred}, Basic Books, N. Y. (2008).

\bibitem{Nagel} E. Nagel, {\it The Structure of Science}, Hackett, Indianapolis (1961); E. Nagel, {\it Teleology revisited: goal-directed processes in biology}, Journ. Philosoph. {\bf 74}, 261 (1977); R. Cummins, {\it The Nature of Psychological Explanation}, MIT Press, Cambridge MA (1983).

\bibitem{functional1} L. H. Hartwell, J. J. Hopfield, S. Leibler, and A. W. Murray, {\it From molecular to modular cell biology}, Nature {\bf 402}, C47 (1999).

\bibitem{functional2} M. Mossio, C. Saborido, and A. Moreno, {\it An organizational account of biological functions}, Br. J. Philos. Sci. {\bf 60}, 813 (2009).

\bibitem{functional3}  G. F. R. Ellis and J. Kopel, {\it The Dynamical Emergence of Biology From Physics: Branching Causation via Biomolecules}, Front. Physiol. {\bf 9}, 1966 (2019).   
	
\bibitem{K-EL} S. Kauffman, {\it Answering Schr\"odingers ``What is Life?''}, Entropy {\bf 22}, 815 (2020).

\bibitem{downward} G. Auletta, G. F. R. Ellis, and L Jaeger, {\it Top-down causation by information control: from a philosophical problem to a scientific research program}, J. Roy. Soc. Interface {\bf 5}, 1159 (2008);  G. Ellis and B. Drossel, {\it How Downwards Causation Occurs in Digital Computers}, Found. Phys. {\bf 49}, 1253 (2019) [arXiv:1908.10186  [quant-ph]]; B. Drossel,   {\it Strong emergence in condensed matter physics},  arXiv:1909.01134  [physics.hist-ph]. 

\bibitem{PT1909} P. Ehrenfest and T. Ehrenfest, {\it Begriffliche Grundlagen der statistischen Auffassung in der Mechanik}, in: Enzyklopidie der mathematischen Wissenschaften mit Einschlflihrer Anwendungen. Band IV, 2. Teil (F. Klein and C. M\"uller (eds.) (1911). Leipzig: Teubner, pp. 3--90. Translated as {\it The conceptual Foundations of the Statistical Approach in Mechanics}, Cornell University Press, N. Y. (1959).

\bibitem{Kinvestigations} S. A. Kauffman, {\it Investigations}, Oxford University Press, N. Y. (2000).

\bibitem{MMMM} M. Mont\'{e}vil and M. Mossio, {\it Biological Organization as a Closure of Constraints}, J. Theor. Biol. {\bf 372}, 179 (2015).

\bibitem{Atkins} P. W. Atkins, {\it The 2nd Law: Energy, Chaos, and Form}, Scientific American Library (1994).

\bibitem{Sch-WIL} E. Schr\"odinger,   {\it What Is Life? The Physical Aspect of the Living Cell}, Cambridge University Press, Cambridge (1944, reissued 2012). Based on lectures delivered under the auspices of the Dublin Institute for Advanced Studies at Trinity College, Dublin, in February 1943.

\bibitem{HSK1} M. Steel, W. Hordijk, and S. A. Kauffman, {\it Dynamics of a birth-death process based on combinatorial innovation}, J. Theor. Biol. {\bf 491}, 110187 (2020) [arXiv:1904.03290 [q-bio.PE]].

\bibitem{OurTAP} M. Cort\^es, S. A. Kauffman, A. R. Liddle, and L. Smolin,
{\it The TAP equation: evaluating combinatorial innovation}, submitted simultaneously.

\end{thebibliography}
\end{document}